%% file: mero_c8m.tex
\documentclass[aps,preprint,nofootinbib,preprintnumbers,eqsecnum,superscriptaddress]{revtex4}

\usepackage[
      colorlinks=true,
      linkcolor=blue,
      urlcolor=blue,
      filecolor=black,
      citecolor=red,
      pdfstartview=FitV,
      pdftitle={},
        pdfauthor={Arpit Das},
        pdfsubject={},
        pdfkeywords={},
        pdfpagemode=None,
        bookmarksopen=true
      ]{hyperref}

\input{preamble.tex}

\begin{document}


\title{Meromorphic CFTs have central charges c = 8$\mathbb{N}$: a proof based on the MLDE approach and Rademacher series}

 \author{Arpit Das}
	\email{arpit.das@ed.ac.uk}
	\affiliation{School of Mathematics, University of Edinburgh, EH9 3FD, U.K.\\}
        \affiliation{Higgs Centre for Theoretical Physics, University of Edinburgh, Edinburgh, EH8 9YL, U.K.}

\begin{abstract}
In this short note, we present a simple and elementary proof that meromorphic conformal field theories (CFTs) have central charges of the form: $c=8N$ with $N\in\mathbb{N}$ (the set of natural numbers) using the modular linear differential equations (MLDEs) approach. We first set up the 1-character MLDE for arbitrary value of the Wronskian index: $\ell$. From this we get the general form of the meromorphic CFT's character. We then study its modular transformations and the asymptotic value of it's Fourier coefficients -- using Rademacher series -- to conclude that odd values of $\ell$ make the character in-admissible implying that the central charge for admissible character has to be a multiple of 8.
\end{abstract}

\vfill

\today
    \hypersetup{linkcolor=black}

\maketitle

\tableofcontents

\section{Introduction}
A rational conformal field theory (RCFT) is a 2D CFT, which has a finite set of holomorphic characters $\chi_i(\tau)$ and a torus-partition function of the form \cite{Anderson:1987ge, Moore:1988qv, Mathur:1988na}:
\be
\mathcal{Z}(\tau,\btau)=\sum_{i,j=0}^{n-1}M_{ij}\,\bchi_i(\btau)\chi_j(\tau) \label{zij}
\ee
where $\tau\in\mathbb{H}=\{z\in\mathbb{C} \, | \, \Im(z)>0\}$ is the modular parameter on the torus, $n$ is the total number of linearly independent characters of some -- typically extended -- chiral algebra $\hat{\frak{g}}$ and $M_{ij}$ denotes the multiplicity matrix where the multiplicity captures the number of primaries sharing the same character. Let us denote the number of primaries by $p$. Usually, we have $p\geq n$ \footnote{This often happens when the chiral algebra includes a Kac-Moody algebra and there is a symmetry in the Dynkin diagram of the corresponding finite-dimensional Lie algebra (see \cite{Das:2023qns, DiFrancesco:1997nk} for more details).}.

A meromorphic CFT is a RCFT which has a single primary field and hence a single-character: the identity character. This means all other fields in the theory are descendants over the identity. In the mathematics literature, meromorphic CFTs are also known as {\it holomorphic vertex operator algebras (VOAs)} (see, for example, definition B.50 in \cite{Rayhaun:2023pgc}). It is well known that the central charge of a meromorphic CFT is a multiple of 8 (see, for instance, \cite{Lin:2021bcp, Dolan:1994st, Lin:2019hks}) however a simple proof of this result is difficult to find in the RCFT literature (see for instance \cite{Goddard:196381} where a similar result has been obtained for lattice RCFTs\footnote{lattice RCFTs are those RCFTs whose chiral algebra is $A, D, $ or, $E$ type affine Lie algebra at level $1$, or, their tensor products} -- using the correspondence between classical self-dual binary linear codes associated with Euclidean even self-dual lattices and chiral bosonic CFTs; for a review see \cite{Dymarsky:2020qom, Dolan:1994st}). The main purpose of this note is to present a simple proof of this ubiquitous fact from the perspective of modular linear differential equations -- for any meromorphic RCFT.

This note is organised as follows. In section \ref{pre2} we review some basic aspects of RCFTs: modular transformations of its characters and their relations to vector-valued modular forms and MLDEs. Readers familiar with the MLDE approach to RCFTs can skip this section and directly move on to section \ref{res3}. In section \ref{res3} we present the main result. First we give the single-character MLDE and from it we get the form of the single-character solution in terms of the {\it Klein-j}, $j(q)$, modular invariant function. We then study its modular transformation properties and use Rademacher sums to analyse the asymptotic behaviour of its Fourier coefficients. This analysis enables us to conclude the even-ness of the Wronskian index, $\ell$, for single-character RCFTs implying that their central charges come in multiples of 8. We conclude this note by a short concluding section \ref{disc4}.

\section{Basics of RCFTs}\label{pre2}
In the section we provide a brief overview of the basic aspects of RCFTs. Let us start by reviewing modular transformations of an RCFT's characters.

\subsection{Modular transformation of RCFT characters}
Since its torus-partition function is of the form given in \eqref{zij} and it is modular invariant:
\begin{align}
    \sZ(\gamma\tau,\gamma\btau) = \sZ(\tau,\btau), \ \ \forall \, \gamma\in\text{SL}(2,\mathbb{Z}) \label{mod_inv_Z}
\end{align}
we get that the holomorphic characters transform into linear combinations among themselves as (see \cite{Zhu:1996} for a rigorous proof and \cite{Harvey:2018rdc, Rayhaun:2023pgc} for more details),
\begin{align}
    \chi_i(\tau) = \sum\limits_{j=0}^{n-1}\rho_{ij}(\gamma)\chi_{j}(\tau), \label{vvmf0}
\end{align}
where $\rho$ is a finite dimensional unitary representation of the modular group SL$(2,\mathbb{Z})$. Characters obeying \eqref{vvmf0} are called vector-valued modular forms (VVMFs) of weight 0. Comparing \eqref{zij} and \eqref{vvmf0} we see that for the partition function to be modular invariant we should have,
\begin{align}
    \rho^\dagger M\rho = M, \label{transf_rho}
\end{align}
We know that SL$(2,\mathbb{Z})$ is generated by the $S=
\begin{pmatrix} 
0 & -1 \\
1 & ~0 \\
\end{pmatrix}$ and $T=
\begin{pmatrix} 
1 & 1 \\
0 & 1 \\
\end{pmatrix}$ matrices and hence one way to determine $\rho$ is to look at its value on these generators, $S:\tau\to-\frac{1}{\tau}$ and $T:\tau\to\tau+1$, subjected to the following modularity constraints (see \cite{Harvey:2018rdc, murty:2016}),
\begin{align}
    (\rho(S))^2 = 1, \qquad (\rho(S)\rho(T))^3 = C, \label{mod_const}
\end{align}
where $C$ is the charge-conjugation matrix satisfying $C^2=\mathds{1}$.

\subsection{MLDEs and RCFTs}
It is known that the characters of an RCFT satisfy a $n^{th}$-order modular linear differential equation (MLDE) of the form \cite{Mathur:1988na, Eguchi:1988wh},
\begin{align}
    \left(\sD^n + \sum\limits_{s=1}^{n-1}\mu_{2s}\phi_{2s}(\tau)\sD^{n-s}\right)\chi(\tau) = 0, \label{mlde0}
\end{align}
where the {\it Serre-Ramanujan} derivative, in $\tau$ (or, in $q\equiv \text{e}(\tau)\equiv e^{2\pi\text{i}\tau}$), denoted $\sD$, is defined as follows. Let 
\begin{align}
\sD_w\equiv \frac{1}{2\pi \text{i}}\frac{\del}{\del\tau}-\frac{w}{12}E_2(\tau) = q\frac{\partial}{\partial q} - \frac{w}{12}E_2(q), \label{Ddef}
\end{align}
be the derivative acting on a modular form of weight $w$. Since $\sD_w$ raises the weight of the form from $w$ to $w+2$, we define:
\begin{align}
\sD^n\equiv \sD_{w+2n-2}\circ \sD_{w+2n-4} \circ \ldots \circ \sD_{w+2}\circ \sD_w. \label{Dndef}
\end{align}
In \eqref{mlde0}, the above definition applies with $w=0$ since characters are weight 0 VVMFs. $\mu_{2s}\in\mathbb{C}$ are arbitrary parameters and $\phi_{2s}$ are meromorphic modular functions of weight $2s$ \footnote{here by meromorphic modular functions we mean modular forms of SL$(2,\mathbb{Z})$ but which are now allowed to have poles in the upper-half plane and hence the term meromorphic. An example would be $\phi_2=\frac{E_6}{E_4}$ where $E_4$ and $E_6$ are the normalised weight 4 and weight 6 Eisenstein series respectively, and $\phi_2$ is a meromorphic modular form of weight 2.}. whose poles are governed by the zeroes of the Wronskian, and whose overall normalisations are specified so that their leading term is unity. Explicitly we have \cite{Mathur:1988na, Mathur:1988gt}:
\be
\mu_{2s}\,\phi_{2s}=(-1)^{s}\frac{W_{n-s}}{W_n}
\label{phidef}
\ee
where:
\be
W_{s}(\tau)\equiv \left| 
\begin{matrix}
\chi_0(\tau) & \chi_1(\tau)& \cdots & \chi_{n-1}(\tau)\\
\vdots & \vdots & \vdots & \vdots \\
\sD_\tau^{s-1}\chi_0(\tau) & \sD_\tau^{s-1} \chi_1(\tau) & \cdots & \sD_\tau^{s-1}\chi_{n-1}(\tau)\\
\sD_\tau^{s+1}\chi_0(\tau) & \sD_\tau^{s+1} \chi_1(\tau) & \cdots & \sD_\tau^{s+1}\chi_{n-1}(\tau)\\
\vdots & \vdots & \vdots & \vdots \\
\sD_\tau^{n}\chi_0(\tau) & \sD_\tau^{n} \chi_1(\tau)
& \cdots & \sD_\tau^{n}\chi_{n-1}(\tau)
\end{matrix}
\right|,
\label{Wronskians}
\ee
where $W\equiv W_n$ is the usual Wronskian. It is easy to see from the above definition \eqref{Wronskians} that $W_{n-1}=\sD W_n$. From \eqref{phidef} we find the following useful relation:
\be
\mu_2\phi_2=-\frac{W_{n-1}}{W_n}=-\sD\log W_n
\label{phitwoprop}
\ee

Note that the modular $T$-transformation on the characters imply the following Fourier expansion, in the variable $q$, also called the $q$-expansion,
\begin{align}
    \chi_i(q) = q^{\alpha_i}\sum\limits_{j\in\mathbb{N}\cup\{0\}}a_{i,j}\,q^j, \label{q_exp}
\end{align}
where the exponents $\alpha_i\equiv -\frac{c}{24}+h_i$ with $c$ being the central charge of the RCFT and $h_i$s being the conformal dimensions of the characters. Also, $h_0=0$ denoting the conformal dimension of the identity character. Now since the solutions of the MLDE -- in the above Frobenius form -- should furnish irreducible representations $\rho$ of the modular group, we have by definition (see \cite{Harvey:2018rdc, Das:2023qns} for details),
\begin{align}
    \rho(T) = \exp{\Big[2\pi \text{i}\,\,{\rm diag}\left(-\sfrac{c}{24},-\sfrac{c}{24}+h_1,\cdots, -\sfrac{c}{24}+h_{n-1}\right)\Big]}, \label{rho_T}
\end{align}

The order of the above MLDE corresponds to the number of linearly independent characters of the RCFT, denoted by $n$. The pole structure of the MLDE is captured by the poles of the coefficient functions $\phi_{2s}$ or the zeros of the Wronskian (see \eqref{phidef}). The zeros of the Wronskian is captured by an integer called the {\em Wronskian index}, denoted by $\ell$. It is defined to be an integer such that $\frac{\ell}{6}$ is the total number of zeroes of $W_n$ (see \cite{Das:2023qns} for more details). 

Note that, $W_n$ is a modular form of weight $n(n-1)$ and at $q\to 0$ its leading behaviour is, $W\sim q^{\sum\limits_{i=0}^{n-1}\alpha_i}$, since $\chi_i(q)\sim q^{\alpha_i}$, applying the valence formula of SL$(2,\mathbb{Z})$ to $W_n$, we get the following all important relation \cite{Mathur:1988na, Harvey:2018rdc, Das:2020wsi},
\begin{align}
    \sum\limits_{i=0}^{n-1}\alpha_i = \frac{n(n-1)}{12} - \frac{\ell}{6}, \label{RR0}
\end{align}
which is also called the {\it Riemann-Roch} relation in RCFTs.

The classification programme of RCFTs based on the above MLDE approach is dubbed the {\it Mathur-Mukhi-Sen} (MMS) classification and it revolves around two parameters of the modular differential equation -- its order and its pole structure, namely: $(n,\ell)$. This programme has been very successful in classifying RCFTs for small values of $(n,\ell)$ and has been pursued by both mathematicians and physicists in more recent times \cite{Naculich:1988xv, Kiritsis:1988kq, Bantay:2005vk, Mason:2007, Mason:2008, Bantay:2007zz, Tuite:2008pt, Bantay:2010uy, Marks:2011, Gannon:2013jua, Kawasetsu:2014, Hampapura:2015cea, Franc:2016, Gaberdiel:2016zke, Hampapura:2016mmz, Arike:2016ana, Tener:2016lcn, Mason:2018, Harvey:2018rdc, Chandra:2018pjq, Chandra:2018ezv, Bae:2018qfh, Bae:2018qym, Mason:c8c16, franc2020classification, Bae:2020xzl, Mukhi:2020gnj, Kaidi:2020ecu, Das:2020wsi, Kaidi:2021ent, Das:2021uvd, Bae:2021mej, Duan:2022ltz, Das:2022slz, Das:2022uoe, Das:2023qns, Duan:2022ltz, Mukhi:2022bte, Rayhaun:2023pgc}

\subsection{Solutions $\to$ admissible solutions}
Once we have set up the MLDE as in \eqref{mlde0}, we seek Frobenius type of solutions to it as given in \eqref{q_exp}. After obtaining these solutions, one has to impose the following constraints,
\begin{enumerate}
    \item Uniqueness of the vacuum -- Corresponding to the identity character's Fourier expansion, we should have $a_{0,0}=1$ which measures the degeneracy of the vacuum
    \item Integrality and positivity of the degeneracies of states -- The Fourier coefficients $a_{0,j}$s, which correspond to the identity character, should be non-negative integers as they capture the state-degeneracies in a {\it Verma module}. For the non-identity characters, the Fourier coefficients are allowed to be positive rationals, since in this case the first term in the $q$-expansion need not begin with unity, as long as these rationals do not grow out of bound and there exists a common denominator -- large but finite -- which when multiplied as an overall factor makes the coefficients $a_{i\neq 0, j}$s integral. In connection to RCFT, this common denominator corresponds to the dimension of the {\it highest weight integral representation} (HWIR) -- with conformal dimension $h_i$ -- of the fully extended chiral algebra of the theory (see \cite{DiFrancesco:1997nk, Hampapura:2015cea, Hampapura:2016mmz, Das:2020wsi} for more details).
    \item Multiplicity of the trivial primary -- The trivial primary correpsonds to the identity character and hence in the torus-partition function in \eqref{zij} we should have $M_{00}=1$.
    \item Positivity and integrality of multiplicities -- Since $M_{ij}$ correspond to the number of primaries that share the same character, we should have $M_{ij}$s as non-negative integers. 
\end{enumerate}
After the solutions to the MLDE satisfy all of the above constraints then they are termed as {\it admissible}.

Note that having admissible solutions doesn't mean having a genuine RCFT. For instance, note that, the following solution to a $(1,6)$ MLDE at central charge 24,
\begin{align}
    j(q) + \sN = q^{-1}(1 + (744+\sN)q + 196884 q^2 + \sO(q^3))
\end{align}
is admissible for $\sN\geq -744$ but only 71 of them correspond to genuine RCFTs \cite{Schellekens:1992db, vanEkeren:2017scl, van2020dimension, vanEkeren:2020rnz, Hohn:2020xfe, Betsumiya:2022avv, Moller:2019tlx, Moller:2021clp}. In the above, $j(q)$ is known as the {\it Klein-j} function which is a modular invariant function.

However, we are not interested in classifying RCFTs in this note. So, we won't discuss the various ways one needs to follow to go from constructing genuine RCFTs from admissible solutions of MLDE. Interested readers can refer to the following articles for more details \cite{Gaberdiel:2016zke, Harvey:2018rdc, Das:2022slz, Das:2022uoe, Mukhi:2022bte, Rayhaun:2023pgc, Duan:2022ltz}. We will only need the admissibilty criterion described above to get to our main result.

\section{$(1,\ell)$ MLDE}\label{res3}
In this section we shall apply the aforementioned constructions to the case of $(1,\ell)$ MLDE. For single-character theory note that, the multiplicity matrix is just, $M=1$ and hence \eqref{transf_rho} becomes $\rho^\dagger\rho = 1$ which implies that the character is allowed to be modular invariant upto an arbitrary phase $\rho=e^{\text{i}\theta}$ with $\theta\in[0,2\pi)$.

Next let us note that for 1-character theory, we have from \eqref{RR0}, that its central charge is related to the Wronskian index as,
\begin{align}
    c=4\ell, \label{imp_rel}
\end{align}
which immediately implies that the central charge of meromorphic theory is a non-negative integer since $\ell\in\mathbb{N}$.

Notice, it is clear from above that, if we can show that meromorphic theories have only even Wronskian indices then it readily follows that their central charges have to be multiples of 8. This is what we intend to show below. To be precise, we will show that odd Wronskian indices lead to in-admissible single-character solutions. Hence, these cannot be realised as characters of a meromorphic RCFT.

Let us start by considering the generic $(1,\ell)$ MLDE, 
\begin{align}
    \left(\sD + \mu_2\phi_2\right)\chi = 0, \label{1_l_gen}
\end{align}
where $\phi_2$ is given as (see Eq.(2.13) of \cite{Das:2023qns}),
\be
\begin{split}
\ell=6r:\quad
\mu_2\,\phi_2 &= E_4^2E_6\sum_{I=1}^r \frac{1}{E_4^3-p_I\Delta}\\
\ell=6r+1:\quad\mu_2\,\phi_2 &= \frac{2E_6}{3E_4}+
\frac{E_4^2}{2E_6}
+E_4^2E_6\sum_{I=1}^{r-1} \frac{1}{E_4^3-p_I\Delta}\\
\ell=6r+2:\quad\mu_2\,\phi_2 &= \frac{E_6}{3E_4}+
E_4^2E_6\sum_{I=1}^r \frac{1}{E_4^3-p_I\Delta}\\
\ell=6r+3:\quad\mu_2\,\phi_2 &= \frac{E_4^2}{2E_6} +E_4^2E_6\sum_{I=1}^r \frac{1}{E_4^3-p_I\Delta}\\
\ell=6r+4:\quad\mu_2\,\phi_2 &= \frac{2E_6}{3E_4}+ E_4^2E_6\sum_{I=1}^r \frac{1}{E_4^3-p_I\Delta}\\
\ell=6r+5:\quad\mu_2\,\phi_2 &= \frac{E_6}{3E_4}+\frac{E_4^2}{2E_6}
+E_4^2E_6\sum_{I=1}^r \frac{1}{E_4^3-p_I\Delta}
\end{split},
\label{phitwobasis}
\ee

Note that, $(n,1)$ MLDE is not allowed by construction \footnote{One quick way to see this is to note that the lowest order zero that the Wronskian can have is of order $\frac{1}{3}$ and not below.}. Thus, we disregard single-character solutions with $\ell=1$ which implies there are no $c=4$ meromorphic theories. 

Now let us consider the MLDE for $\ell=6r+1$ with $r=1$ and obtain its solution. It is easier to work in the $j$-space and so we will transform the $\tau$-space MLDE to the $j$-space MLDE using the following identities,
\begin{equation}
\begin{split}
    &j = \frac{E_4^3}{\Delta}, \qquad (j-1728) = \frac{E_6^2}{\Delta}, \\
    &\sD\chi = -j\frac{E_6}{E_4}\partial_j\chi,
\end{split}\label{iden_j}
\end{equation}
where $\Delta$ is the modular discriminant and is a cusp form of weight 12 for the modular group. Sunc transformations of MLDEs between the $\tau$-space and the $j$-space have been recently studied in \cite{Das:2023qns}. We get,
\begin{align}
    \partial_j\chi - \left(\frac{2}{3j} + \frac{1}{2(j-1728)} + \sum\limits_{I=1}^{r-1}\frac{1}{j-p_I}\right)\chi = 0, \label{j_1_0}
\end{align}
The solution to the above equation can be readily given as,
\begin{align}
    \chi(j) = j^{\frac{2}{3}}(j-1728)^{\frac{1}{2}}\prod\limits_{I=1}^{r-1}(j-p_I), \label{j_1_0_soln}
\end{align}
where the integration constant is fixed by demanding that the leading term in the $q$-expansion of $\chi$ is unity. Similarly, one can immediately write down the solutions for the other two remaining odd cases. For $\ell=6r+3$ we get,
\begin{align}
    \chi(j) = (j-1728)^{\frac{1}{2}}\prod\limits_{I=1}^{r}(j-p_I), \label{j_3_0_soln}
\end{align}
and for $\ell=6r+5$ we get,
\begin{align}
    \chi(j) = j^{\frac{1}{3}}(j-1728)^{\frac{1}{2}}\prod\limits_{I=1}^{r}(j-p_I). \label{j_5_0_soln}
\end{align}

Now let us consider all the individual multiplying factors listed in the above solutions. Note, $j(q)$ is admissible -- as it happens to be the character of the meromorphic theory $E_{8,1}^3$ -- and hence $j(q)^{\frac{1}{3}}$ and $j(q)^{\frac{2}{3}}$ are also admissible -- which are the characters of the meromorphic theories $E_{8,1}$ and $E_{8,1}^2$ or, $\mathcal{E}_1[D_{16,1}]$ respectively\footnote{where the notation $\mathcal{E}_1[\hat{\mathfrak{g}}]$ denotes the 1-character extension of the $n$-character meromorphic theory with chiral algebra $\hat{\mathfrak{g}}$ (see \cite{Mukhi:2022bte, Das:2022uoe, Rayhaun:2023pgc} for more details).} (see \cite{Rademacher:1939fou, Schellekens:1992db, Das:2022uoe, Mukhi:2022bte, Rayhaun:2023pgc} for more details). However, the factor $(j-1728)^\frac{1}{2}$ is in-admissible and has all negative Fourier coefficients except the first term which is unity \cite{Chandra:2018pjq}. This rules out any admissible solutions at $\ell=3$ or $c=12$.

In spite of this, it might still happen that considering enough $(j-p_I)$ factors might turn the overall in-admissible solution into an admissible solution. So, at higher central charges, odd $\ell$ values still remains a possibility.

\subsection{Elementary proof}
To remedy this, we need to analyse the asymptotic behaviour of the above solutions. For this let us employ the {\it Rademacher sums} analysis as outlined in appendix A of \cite{Chandra:2018pjq} (see also \cite{Cheng:2012qc} for more details). Using Eq.(A.4) of \cite{Chandra:2018pjq} for single-character solution we get,
\begin{align}
    &a_0(j) = \rho(S)_{00}^{-1} \, a_{0}(0)\exp\left\{4\pi\sqrt{\frac{\ell}{6}\left(j-\frac{\ell}{6}\right)}\right\} + \text{(sub-leading)}, \nonumber\\
    &a_0(j) = \rho(S)_{00}^{-1} \, \exp\left\{4\pi\sqrt{\frac{\ell}{6}\left(j-\frac{\ell}{6}\right)}\right\} + \text{(sub-leading)}, \label{Rade0}
\end{align}
where $j$ denotes the order in the $q$-expansion in \eqref{q_exp}. The above holds in the limit of $j\to\infty$. In above, going to the second-line we have used $a_0(0)=1$. So, $\text{sgn}(\rho(S)_{00}^{-1})=\text{sgn}(a_0(j))$ for $j\to\infty$. Let us determine $\text{sgn}(\rho(S)_{00}^{-1})$ next.

Since $c=4\ell$, we have $\rho(T) = \text{e}(-c/24) = e^{-\text{i}\pi\frac{\ell}{3}}$. From the first modularity constraint in \eqref{mod_const} we have, $\rho(S)^2 = C = e^{\text{i}m\pi}$. This implies,
\begin{equation}
    \rho(S) = e^{m\text{i}\pi/2}. \label{c1}
\end{equation}
The second modularity constraint  in \eqref{mod_const} we have $(\rho(S)\rho(T))^3 = C = e^{\text{i}m\pi}$. This implies,
\begin{align}
    \exp\left[\text{i}\pi\left(\frac{m}{2}-\ell\right)\right] = 1 = \text{e}(k), \label{c2}
\end{align}
where $m, k$ are integers. Thus, we get from above:
\begin{equation}
\begin{split}
    & \frac{m}{2} - \ell = 2k, \\
    \text{implying,} \, \, & \ell = \frac{m}{2} - 2k, \\
    \text{implying,} \, \, & m>4k \qquad \text{ and, } m\in 2\mathbb{N}
\end{split}, \label{res}    
\end{equation}
since $\ell\in\mathbb{N}$.

Now consider odd $\ell$ which implies $m=4v+2$ (from \eqref{res}), with $v$ being an integer. Then, we get: $\rho(S) = e^{\text{i}\pi}=-1$ which implies $\text{sgn}(\rho(S_{00})^{-1})=-1=\text{sgn}(a_0(j))$ implying that odd $\ell$ results in in-admissible characters solutions to the MLDE. Note that this in-admissibility essentially arises from the fact that $(j-1728)^\frac{1}{2}$ has all Fourier coefficients negative except the first term and this spoils the admissibility as we go to higher and higher orders in the $q$-expansion or, as $j\to\infty$ in \eqref{q_exp}.

For the sake of completeness, let us consider even $\ell$, which implies $m=4v$ (from \eqref{res}), with $v\in\mathbb{Z}$. Then, we get: $\rho(S)=\text{e}(v)=1$. Thus, $a_0(j)>0$ for $j\to\infty$ in \eqref{q_exp}.

Thus, we see from above that single-character solutions to $(1,\ell)$ MLDE with odd $\ell$ lead to in-admissible solutions implying that for admissible solutions $\ell$ must be even. So, the space of genuine RCFTs must have even $\ell$ implying that their charges will be $c = 4\ell = 8N$ with $N\in\mathbb{N}$. With this we can write down the ansatz for the most general single-character admissible solution as,
\begin{align}
    \chi(q) = j^{\frac{N}{3}-s}(j^s + a_1\,j^{s-1}+ a_2 \, j^{s-2} + \ldots\ldots + a_s),
\label{gen_j}
\end{align}
where $s := \left \lfloor \frac{N}{3} \right \rfloor$. This ansatz is also given in Eq.(2.2) of \cite{Das:2022slz}.

\section{Conclusion}\label{disc4}
In this short note, we presented an elementary proof of the fact that central charges of meromorphic CFTs are multiples of 8. The approach we took in proving this was studying the MLDEs that single-character theories' satisfy and by observing their solution space. We then analysed the asymptotic behaviour of the Fourier coefficients of the single-character solutions using Rademacher sums which led to the conclusion that odd $\ell$ solutions are in-admissible. Thus, admissible solutions and hence genuine meromorphic CFTs will have even $\ell$ implying that their central charges will be multiples of 8.

Such analysis of the values of the Wronskian index was done using MLDEs for the two-character case in \cite{Das:2023qns} where it led to the conclusion that the Wronskian index should be even for all two-character admissible solutions too. There is an older result on the even-ness of $\ell$ for the two-character case in \cite{Naculich:1988xv} where this fact was proven using monodromy arguments of the modular $S$ and $T$ matrices. In an ongoing work, we are trying to extend this analysis to three or more character-solutions and MLDEs. We are hopeful that by studying the relevant MLDEs and their solution spaces -- across all the orbifold points in the fundamental domain of SL$(2,\mathbb{Z})$ -- we should be able to re-derive established results like: $\ell=3\mathbb{N}$ for the three-character case, $\ell=2\mathbb{N}$ for the four-character case, $\ell=\mathbb{N} \setminus \{1\}$ for the five-character case, etc., as given in \cite{Kaidi:2021ent}; and also extend upon these results.

\section*{Acknowledgements}

We would like to thank Chethan N. Gowdigere, Sunil Mukhi, Jagannath Santara, Jishu Das and Naveen Balaji Umasankar for useful discussions on modular forms and MLDEs. We are also immensely grateful to Sigma Samhita and Shaymin for their constant support and love. AD is supported by the STFC Consolidated Grant ST/T000600/1 -- ``Particle Theory at the Higgs Centre''.

\bibliographystyle{utphys}

\bibliography{mero_c8m}

\end{document}

%% file: preamble.tex
\usepackage{color}

\newcommand{\Comment}[1]{{}}

\usepackage{threeparttable}

\usepackage[normalem]{ulem}
\usepackage{amsmath}
\usepackage{enumerate}
\usepackage{amsfonts}
\usepackage{yfonts}
\usepackage{array,xcolor,graphicx}

\usepackage{subfigure}
\usepackage{psfrag}

\usepackage{epsfig}
\usepackage{float}
\usepackage{dsfont}
\usepackage{graphicx}
\usepackage{cancel}
\usepackage{mathrsfs}
\usepackage{amssymb}
\usepackage{amsfonts}
\usepackage{amsmath}
\usepackage{slashed}
\usepackage{feynmp}
\usepackage{tikz-feynman}
\tikzfeynmanset{compat=1.1.0}

\usepackage{graphicx}
\usepackage{bm}

\def\del{\partial}

\def\({\left(}
\def\){\right)}
\def\[{\left[}
\def\]{\right]}
\def\<{\langle}
\def\>{\rangle}

\newcommand{\bmat}{\begin{bmatrix}}
\newcommand{\emat}{\end{bmatrix}}

\newcommand\sfrac[2]{\ensuremath{\frac{#1}{#2}}}

\newcommand{\be}{\begin{equation}}
\newcommand{\ee}{\end{equation}}
\newcommand{\bea}{\begin{eqnarray}}
\newcommand{\eea}{\end{eqnarray}}
\newcommand{\bwt}{\begin{widetext}}
\newcommand{\ewt}{\end{widetext}}

\newcommand{\bi}{\begin{itemize}}
\newcommand{\ei}{\end{itemize}}
\newcommand{\ben}{\begin{enumerate}}
\newcommand{\een}{\end{enumerate}}
\newcommand{\bca}{\begin{cases}}
\newcommand{\eca}{\end{cases}}
\newcommand{\bln}{\begin{align}}
\newcommand{\eln}{\end{align}}
\newcommand{\bst}{\begin{split}}
\newcommand{\est}{\end{split}}

\newcommand\sD{{\ensuremath{{\mathcal D}}}}

\newcommand\sN{{\ensuremath{{\mathcal N}}}}
\newcommand\sO{{\ensuremath{{\mathcal O}}}}

\newcommand\sZ{{\ensuremath{{\mathcal Z}}}}

\renewcommand{\Im}{\textrm{Im}\,}

\newcommand{\bchi}{\bar{\chi}}
\newcommand{\btau}{\bar{\tau}}